\documentclass[%
 aip,
 amsmath,amssymb,
 reprint,%
]{revtex4-1}

\usepackage{graphicx}% Include figure files
\usepackage{dcolumn}% Align table columns on decimal point
\usepackage{bm}% bold math
\usepackage[caption=false]{subfig}
\usepackage{hyperref}
\usepackage[utf8]{inputenc}
\usepackage[T1]{fontenc}
\usepackage{mathptmx}
\usepackage{xcolor}

\draft % marks overfull lines with a black rule on the right

\begin{document}

%\preprint{}

\title{Spectroscopy of phase transitions for  multiagent systems} %Title of paper

\author{Niccol\`o Zagli}
 \email{n.zagli18@imperial.ac.uk}
 \affiliation{Department of Mathematics, Imperial College London, London, SW7 2AZ, UK}
 \affiliation{Department of Mathematics and Statistics, University of Reading, Reading, RG6 6AX, UK}
 \affiliation{Centre for the Mathematics of Planet Earth, University of Reading, Reading, RG6 6AX, UK}

\author{Valerio Lucarini}
\affiliation{Department of Mathematics and Statistics, University of Reading, Reading, RG6 6AX, UK}
 \affiliation{Centre for the Mathematics of Planet Earth, University of Reading, Reading, RG6 6AX, UK}
 
 \author{Grigorios A. Pavliotis}
 \affiliation{Department of Mathematics, Imperial College London, London, SW7 2AZ, UK}

\date{\today}

\begin{abstract}
In this paper we study phase transitions for weakly interacting multiagent systems.  By investigating the linear response of a system composed of a finite number of agents,  we are able to probe the emergence in the thermodynamic limit of a singular behaviour of the susceptibility. We find clear evidence of the loss of analyticity due to a pole crossing the real axis of frequencies. Such behaviour has a  degree of universality, as it does not depend on either the applied forcing nor on the considered observable. We present results relevant for both equilibrium and nonequilibrium phase transitions by studying the Desai-Zwanzig and Bonilla-Casado-Morillo models.
\end{abstract}


\maketitle 

\section*{}
\textbf{Multiagent models feature in a very vast range of applications in natural sciences, social sciences, and engineering. We study here the Desai-Zwanzig and Bonilla-Casado-Morillo models, which are paradigmatic for equilibrium and nonequilibrium conditions, respectively. Phase transitions result from the coordination between the individual agents, and are associated with the divergence of the linear response of the system. The occurrence of phase transitions  is universal: it does not depend on the acting forcing, and can be detected by looking at virtually any observable of the system. We showcase here how response theory is capable of providing a useful angle for understanding the universal properties of phase transitions in complex systems.
}

\section{\label{sec:Introduction} Introduction}
Agent based models are regularly employed to model various phenomena in the natural sciences, social sciences and engineering~\cite{NaldiParentiToscani,toscani2014}. 
Multiagent systems are used to model diverse phenomena such as cooperation~\cite{Dawson}, synchronisation ~\cite{Kuramoto}, systemic risk~\cite{risk} and consensus formation ~\cite{HasgelmannKrause,HasgNumerics}. They are fundamental in developing algorithms for sampling and optimization~\cite{reich2020}  and they have also been used for the management of natural hazard~\cite{Simmonds2019} and climate change impact~\cite{Geisendort2017}. 

Multiagent systems can often exhibit abrupt changes in their behaviour, often corresponding to critical transitions that occur when a parameter, e.g. interaction strength or temperature, passes a certain threshold. Such transitions are often associated to cataclysmic events such climate change, market crashes etc~\cite{scheffer2009critical,sornette2006}. The importance of developing tools for predicting critical transitions has long been recognized. One of the main tools used in order to develop early warning signals for critical transitions is that of linear response theory. % \cite{Chekroun2014,Tantet2018}.

Following the seminal contribution by Kubo \cite{Kubo.1966}, linear response theory represents a very powerful framework for studying the properties of statistical mechanical systems by investigating how they respond to external perturbations \cite{ReviewLinearResponse,Baiesi2013,RespTheoryVulp}. %When considering near-equilibrium conditions, the fluctuation-dissipation theorem establishes a connection between exogenous perturbations and endogenous dynamics, thus giving the opportunity of predicting the response of the system by looking at the natural, unperturbed  fluctuations \cite{Kubo.1966}. %investigating the system by probing it with small external forcings. 
Linear response theory has been successfully applied to classic problems of solid-state physics and optics~\cite{Lucarini2005} as well as plasma physics and stellar dynamics \cite[Ch.5]{BinneyTremaine2008}; see some examples of application of the theory in  %Furthermore, it has proven itself as a fruitful approach to a wide range of systems, spanning both 
both equilibrium and nonequilibrium systems~\cite{Leith1975,North1993,Ottinger2005,Lucarini2017,Cessac2019,Gottwald2020}. %Linear response has also been employed to predict the response of the climate system to external  forcings~\cite{Leith1975,North1993,gritsun2008b,Lucarini2017,Bodai2020,Lembo2020}. 
Rigorous mathematical foundations for linear response theory have been provided for the case of Axiom A systems \cite{ruelle_nonequilibrium_1998,ruelle_review_2009} (see e.g. \cite{B14} for further developments in the context of deterministic systems) and for diffusion processes, both in finite and in infinite dimensions \cite{DemboDeuschel2010, HairerMajda2010}; see also the interesting contributions \cite{Wormell2019} that bridges the deterministic and the stochastic viewpoints.%An up-to-date perspective on recent developments in linear response theory can be found in \cite{Gottwald2020}. 

Critical transitions arise when the spectral gap of the transfer operator of the unperturbed system shrinks to zero \cite{liverani2006,Chekroun2014,Lucarini2016,Tantet2018} as the  Ruelle-Pollicott poles ~\cite{Pollicott1985,Ruelle1986}, touch the real axis.  Near criticality, the negative feedbacks of the system become increasingly ineffective, resulting in arbitrarily large, usually non-Gaussian, fluctuations and a divergence of correlation properties of the system~\cite{Dawson, Shiino1987,Delgadino2021}. 

In the thermodynamic limit, multiagent system can also undergo a qualitative change of their properties through a different mathematical mechanism, namely phase transitions~\cite{Dawson, Shiino1987}, defined as exchange of stability of nonunique stationary distributions as the parameters of the systems vary; see a detailed analysis in \cite{Carrillo:2020aa}. 
%\subsection{\label{sec:level1}Breakdown of the Linear Response in Agent Based Models}% and critical phenomena}

In a previous paper \cite{FirstPaper} we derived  linear response formulas for a system of weakly interacting diffusions described by an $N-$particle Fokker-Planck equation and have explicitly identified two qualitatively different scenarios for the breakdown of the linear response, associated with the previously mentioned critical transitions and phase transitions. We focus here on the latter case. 
%In many cases, linear response theory breaks down at the critical transition since there is a one-to-one correspondence between the radius of expansion of linear response and the spectral gap of the transfer (Fokker-Planck) operator of the unperturbed system  \cite{liverani2006,Lucarini2016},  as the resonances of the system, the Ruelle-Pollicott poles ~\cite{Pollicott1985,Ruelle1986,Chekroun2014}, touch the real axis. 
%It is by now well understood that 
%Critical transitions arise when the the spectral gap of the transfer operator of the unperturbed system shrinks to zero \cite{liverani2006,Chekroun2014,Lucarini2016} as as the  Ruelle-Pollicott poles ~\cite{Pollicott1985,Ruelle1986}, touch the real axis.  Near criticality, the negative feedbacks of the system become increasingly ineffective, resulting in arbitrarily large, usually non-Gaussian, fluctuations and a divergence of correlation properties of the system~\cite{Dawson, Shiino1987,MGRGGP2020}. %This critical slowing down of the system has often been  proposed as an early warning signal for tipping points ~\cite{SlowingDown,Kuehn2,Scheffer}. %In the systems treated here, the previous statement applies to the  transfer operator acting on small perturbations with respect to the reference invariant measures (keep in mind that the McKean-Vlasov equation is nonlinear).  
 %Furthermore, in the proximity of a critical transition one encounters rough dependence of the properties of the system on its parameters~\cite{Chekroun2014,Tantet2018}.
%On the other hand, 
Phase transitions %, in contrast to critical/noise-induced transitions, 
are a genuine thermodynamic phenomenon, where the divergence of the response stems from the coordination taking place, in suitable conditions, because of the coupling between the infinite number of agents composing the total system. %Phase transitions don't conform to the classic critical transition framework as described above. 
The coupling among the subsystems results in a memory effect that leads to obtaining the macroscopic response function of the system as a renormalised version of its  microscopic counterpart~\cite{FirstPaper}, with formal similarities  with the  well-known Clausius-Mossotti relation ~\cite{Lucarini2005,Jackson1975,Talebian2013}. % that connects the macroscopic polarizability of a material and the microscopic polarizability of its microscopic components. 
The role of memory in determining criticality due to endogenous processes has been emphasised in \cite{Sornette2003c,sornette2006}. % and mirrors phenomena %This result, which formally resembles , %in renormalisation  process of microscopic laws into macroscopic observable features 
%Despite formal differences,  is in the same spirit as what has been observed in systems close to a 
%observed in the regime of self organised criticality~\cite{BakSOC,EarthquakesSOC} for models of social interactions~\cite{Sornette2006} and in geophysical sciences~\cite{SornetteEarthquakes}. %We also remark that this phenomenon can be interpreted~\cite{FirstPaper} as a generalisation  of the  
%As opposed to critical transitions, the spectral gap of the transfer operator associated to the steady state regime of the system ~\cite{FRANKResponse,FirstPaper}, does not vanish at the phase transition.  
The link between phase transitions and slow decay of correlations for interacting particle systems is well established, see. e.g.~\cite{yoshida_2003}. % and cannot resort to the early warning indicators mentioned above.

\subsection{\label{sec:letter}This Paper}
%We investigate the phase transitions of such systems using % The goal of this letter is to explore such phase transitions in the thermodynamic limit of multiagent systems using rather classical mathematical techniques based on 
%dispersion relations that have been developed for investigating the optical properties of atoms, molecules, fluids, and condensed matter \cite{Lucarini2005}. 
In this paper we focus in much greater detail on the relationship between the occurrence of phase transitions and the non-analyticity of the susceptibility of the system describing the frequency-dependent response of an observable to a given perturbation in the upper complex frequency plane. The singularity manifests itself as a pole that crosses the real axis of the frequency variable. We use a formalism that mirrors spectroscopic techniques that are  used for investigating the frequency dependence of the optical properties of materials \cite{Lucarini2005}.  By studying how the real and imaginary part of the susceptibility of the systems depend on the number of agents,  we are able to predict the position of the pole and the associated residue, which describe  the emergence of the singularity in the thermodynamic limit. We verify that the position of the pole depends on the considered model, but, instead, that for a given model the loss of analyticity depends neither on the choice of the observable, nor on the applied perturbation, and is, in this sense, an universal feature of the system.%(except degenerate, non-typical cases). and is in this sense universal in nature \cite{FirstPaper}. 
 Our numerical investigations are performed on the Desai-Zwanzig (DZ)~\cite{DesaiZwanzig} and the Bonilla-Casado-Morillo (BCM)~\cite{Bonilla1987} models. The DZ model %, that is closely related to the classical Curie-Weiss model,  
 exhibits a paradigmatic example of an equilibrium order-disorder phase transition, analogous to the Ising ferromagnetic transition \cite{Shiino1987,Dawson}, while the BCM model describes an out-of-equilibrium synchronisation transition of an infinite collection of coupled nonlinear oscillators. As the transition point is crossed, the order parameter (magnetization)
 %$\langle \mathbf{x} \rangle$  
 acquires a non vanishing constant value for the DZ model and is periodically  oscillating for the BCM model. 
%present letter, we will show how linear response theory provides a relevant framework to address the critical behaviour of  at the phase transition. As a matter of fact, the linear response of the system is not related to its correlation properties and depends on the spectrum of a suitably defined operator that takes into account the coupling between the subsystems. At the phase transition, linear response theory breaks down, with the macroscopic renormalised response function developing non-analytical  features, thus modifying the classic dispersion relations~\cite{FirstPaper}.

\section{\label{sec:citeref}The general framework}

We investigate a system composed of $N$ exchangeable interacting $M$ dimensional sub-systems whose dynamics is determined by the following It\^{o} stochastic differential equations (SDEs)
\begin{equation}\label{eq1}
\mathrm{d}\mathbf{x}^{k}=\mathbf{F}_{\alpha}(\mathbf{x}^k) \mathrm{d}t-\frac{\theta}{N}\sum_{l=1}^N \mathbf{\nabla}\mathcal{U}\left(\mathbf{x}^k-\mathbf{x}^l\right)\mathrm{d}t+\sigma \mathbf{S}(\{\mathbf{x}^k\}) \mathrm{d}\mathbf{W}, 
\end{equation}
where $\mathbf{x}^k \in \mathbb{R}^M$ and $k = 1,\ldots,N $.
$\mathbf{F_\alpha}:\mathbb{R}^M\rightarrow  \mathbb{R}^M$ is a smooth vector field, possibly depending on a parameter $\alpha$, and $\mathbf{W}$ denotes a standard $P-$dimensional Brownian motion; $\mathbf{S}:\mathbb{R}^M\rightarrow  \mathbb{R}^{M\times P}$ is the volatility matrix and  the parameter $\sigma>0$ controls the strength of the stochastic forcing, i.e. plays the role of the temperature. 
We consider a fully coupled system given by the quadratic (Curie-Weiss) interaction potential $\mathcal{U}\left(\mathbf{y} \right)=\frac{|\mathbf{y}|^2}{2}$. In this case, the order parameter is known and it is given by the first moment/magnetization. 
%In this Letter we will consider a quadratic interaction potential  $\mathcal{U}\left(\mathbf{y} \right)=\frac{|\mathbf{y}|^2}{2}$. 
%In this case, the $N$ sub-systems undergo an all-to-all coupling through the Laplacian matrix of the underlying graph. For quadratic (Curie-Weiss) interactions, the order parameter is known and it is given by the first moment/magnetization. 

%More general interaction potentials will be considered in future work. We expect that the results presented in this Letter are valid for a broad class of interaction potentials. 
%We remark that the below results refer to a critical mechanism that do not depend on the choice of the interaction potential .  
The coefficient $\theta$ modulates the intensity of the coupling, which attempts at synchronising all systems by attracting them to the center of mass. %$\frac{1}{N}\sum_{k=1}^N \mathbf{x}^k$.
In the thermodynamic limit $N \rightarrow \infty$ the one-particle distribution function converges to the distribution $\rho(\mathbf{x},t)$ that satisfies a nonlinear and nonlocal Fokker-Planck equation~\cite{Dawson,Snitz,oelschlager1984,gartner}
\begin{equation}\label{eq : infinite system}
\begin{split}
\partial_t \rho(\mathbf{x},t) =&-\nabla \cdot \left[\rho(\mathbf{x},t) \left(\mathbf{F}_\alpha(\mathbf{x})+\theta \left(\langle\mathbf{x}\rangle(t)-\mathbf{x}\right) \right) \right] \\
&+\frac{\sigma^2}{2}  \nabla \cdot \nabla \cdot (\mathbf{D} \rho(\mathbf{x},t)),
\end{split}
\end{equation}
where $\mathbf{D}  = \mathbf{S}\mathbf{S}^T$. This mean field Partial Differential Equation might support multiple coexisting stationary measures at low temperatures/large interaction strengths. In particular, in a conservative system described by a confining potential $\mathbf{F_\alpha}(\mathbf{y})=-\nabla V_\alpha(\mathbf{y})$ with additive noise such that $\mathbf{S}$ is the identity matrix, stationary solutions of Eqn.~\ref{eq : infinite system} correspond to local minima of $V_{\alpha}(\mathbf{y})$. In this case, the thermodynamic limit~\eqref{eq : infinite system} can be written in the standard form as $\lim_{N \rightarrow +\infty} -\frac{1}{N} \log Z_N = \inf F(\rho)$, where $Z_N$ denotes the partition function of the $N-$particle system and $F(\rho)$ denotes the free energy of mean field system~\cite{Helffer2002}. A stationary state is characterised by the order parameter $\langle \mathbf{x} \rangle_0$ and the associated stationary distribution $\rho_0(\mathbf{x})$. 
%In particular, it is well known that the non-degeneracy of the spectral gap of the Fokker-Planck operator results in the absence of phase transitions and in the exponentially fast decay of correlations~\cite{yoshida_2003}.

%Bifurcations of stationary solutions as the parameters of the system vary represent critical points for the system.
%The goal of this Letter is to investigate how a stationary reference state $\rho_0(\mathbf{x})$ responds to exogenous perturbations as it gets closer to a critical point.
We now perturb  the stationary state by setting $\mathbf{F}_\alpha(\mathbf{x})\rightarrow \mathbf{F}_\alpha(\mathbf{x})+\varepsilon \mathbf{X}(\mathbf{x})T(t)$ and we study the response of the system by expanding the distribution function as $\rho(\mathbf{x},t)=\rho_0(\mathbf{x})+\varepsilon \rho_1(\mathbf{x},t) + O(\varepsilon^2)$. Following a tedious calculation reported in \cite{FirstPaper}, the response of the order parameter in the frequency domain is written in terms of a macroscopic (or renormalised) susceptibility $\tilde{\Gamma}_i(\omega)$ as (Repeated indices are summed):
\begin{equation}\label{eq : renormalised response}
\langle x_i \rangle_1(\omega)=  P_{ij}^{-1}(\omega)\Gamma_j(\omega) T(\omega)\equiv \tilde{
\Gamma}_i(\omega)T(\omega) 
\end{equation}where $P_{ij}(\omega) = \delta_{ij} - \theta\mathcal
Y_{ij}(\omega)$ and the susceptibilities $\Gamma_i(\omega)$and $\mathcal
Y_{ij}(\omega)$ are respectively the Fourier Transform of the microscopic response functions that can be written as correlation functions in the unperturbed state as~\cite{FirstPaper} 
\begin{align}
G_i(\tau) &= - \Theta(\tau)\left\langle \frac{\nabla \cdot \left (\rho_0(\mathbf{y}) \mathbf{X}(\mathbf{y})\right)}{\rho_0(\mathbf{y})}\exp\left(\mathcal{L}^+_{\langle \mathbf{x} \rangle_0}\tau\right) y_i \right\rangle_0\\
Y_{ij} (\tau)&=- \Theta(\tau)\left\langle  \partial_{y_j} \log\rho_0(\mathbf{y})\exp\left(\mathcal{L}^+_{\langle \mathbf{x} \rangle_0}\tau\right) y_i \right\rangle_0
\end{align}
where $\mathcal{L}^+_{\langle \mathbf{x} \rangle_0}$represents the adjoint of  $\mathcal{L}_{\langle \mathbf{x} \rangle_0}$ and  $\langle \cdot \rangle_0$ is the expectation value on the unperturbed state $\rho_0(\mathbf{x})$. The Fokker-Planck operator $\mathcal{L}_{\langle \mathbf{x} \rangle_0}$ appears on the right hand side of~\eqref{eq : infinite system}  (evaluated at the stationary state $\rho_0(\mathbf{x})$) and its adjoint $\mathcal{L}^+_{\langle \mathbf{x} \rangle_0}$ can be interpreted as the generator of the  Koopman operator of the stationary dynamics  ~\cite{FRANKResponse,FirstPaper}. As such, correlation properties of the system are related to the spectrum of $\mathcal{L}^+_{\langle \mathbf{x} \rangle_0}$~\cite{ChekrounJSPI,Chekroun2014,lasota}. The renormalisation %process 
of the susceptibility %, corresponding to the left-multiplication times the matrix $P_{ij}^{-1}$
derives from the coupling among the subsystems; note that %In the absence of coupling, the matrix $P_{ij}$ is simply the identity matrix and $\tilde{\Gamma}_i \equiv \Gamma_i$. 
%Equation \ref{eq : renormalised response} shows that, in general, 
$\tilde{\Gamma}_i(\omega)$ inherits the poles of both  $\Gamma_i(\omega)$ and  of the matrix $P^{-1}_{ij}(\omega)$. Away from criticality, both the microscopic and macroscopic susceptibilities are analytic in the upper half of the $\omega$  complex plane. 
\begin{figure*}
    \subfloat[]{%
     \includegraphics[trim= 0.2cm 0cm 1.2cm 1.3cm,clip=true, scale=0.5]{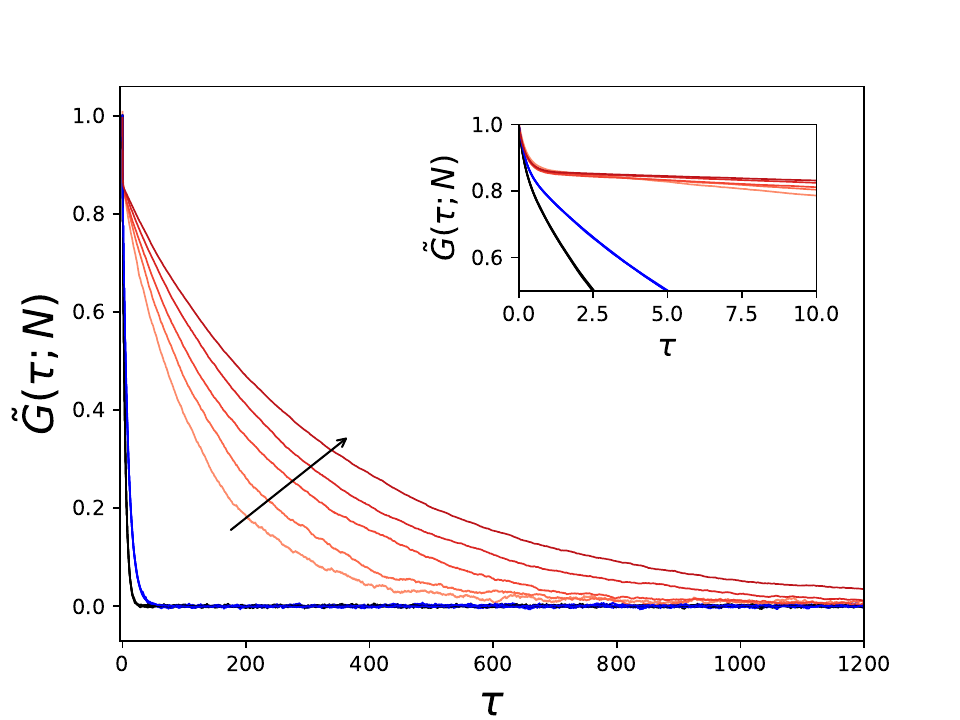}%trim=0.15cm 0.1cm 1cm 1.2cm, clip=true,scale=0.53
     \label{fig: subfig 1a}
    }
    \qquad \qquad
    \subfloat[]{%
     \includegraphics[trim= 0.2cm 0cm 1.2cm 1.3cm,clip=true, scale=0.5]{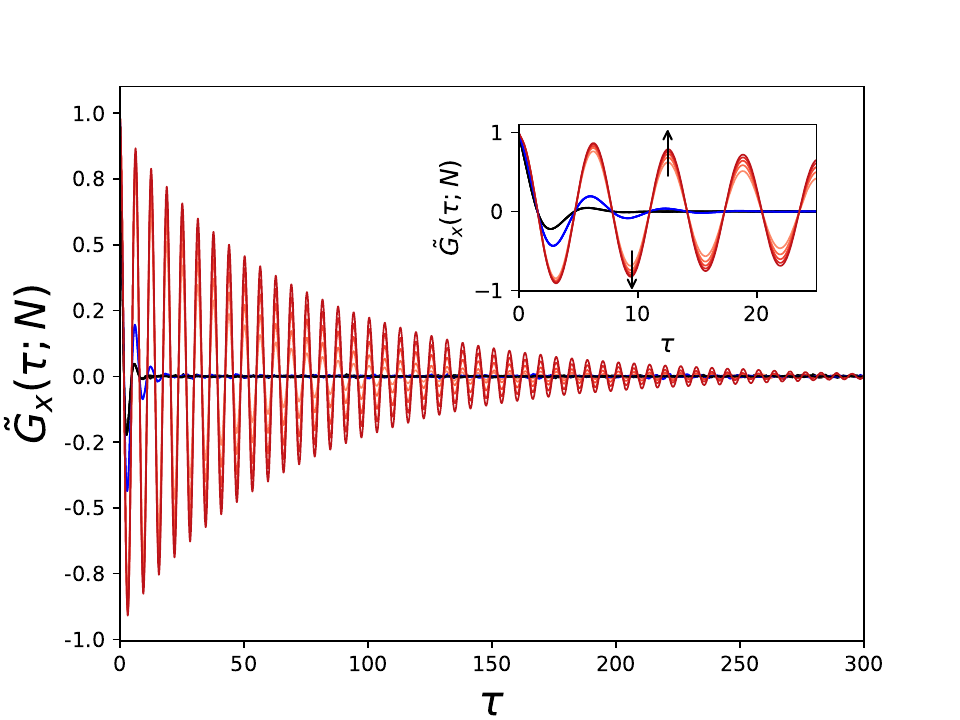}
    \label{fig: subfig 1b}
    }
    \caption{Renormalised response functions as a function of time $\tau$. Panel \protect\subref{fig: subfig 1a}: response $\tilde{G}(\tau;N)$ for the one dimensional order parameter of the DZ model. Panel \protect\subref{fig: subfig 1b}:  response $\tilde{G}_x(\tau;N)$ of the first component of the bi-dimensional order parameter for the BCM model. Black and blue lines correspond to non critical values of the strength of the noise $\sigma$. Red lines correspond to response functions at the transition point. For each value of $\sigma$, there are five lines corresponding to different values of $N$, namely $N = 2^k \times 10^3$ with $k = 1, \dots, 5$. The arrows and the colour gradient indicate the direction of increasing $N$.}
    \label{fig: Response in time}
\end{figure*}
\begin{figure*}
    \subfloat[]{%
    \includegraphics[trim= 0.15cm 0cm 1.2cm 1.3cm,clip=true, scale=0.5]{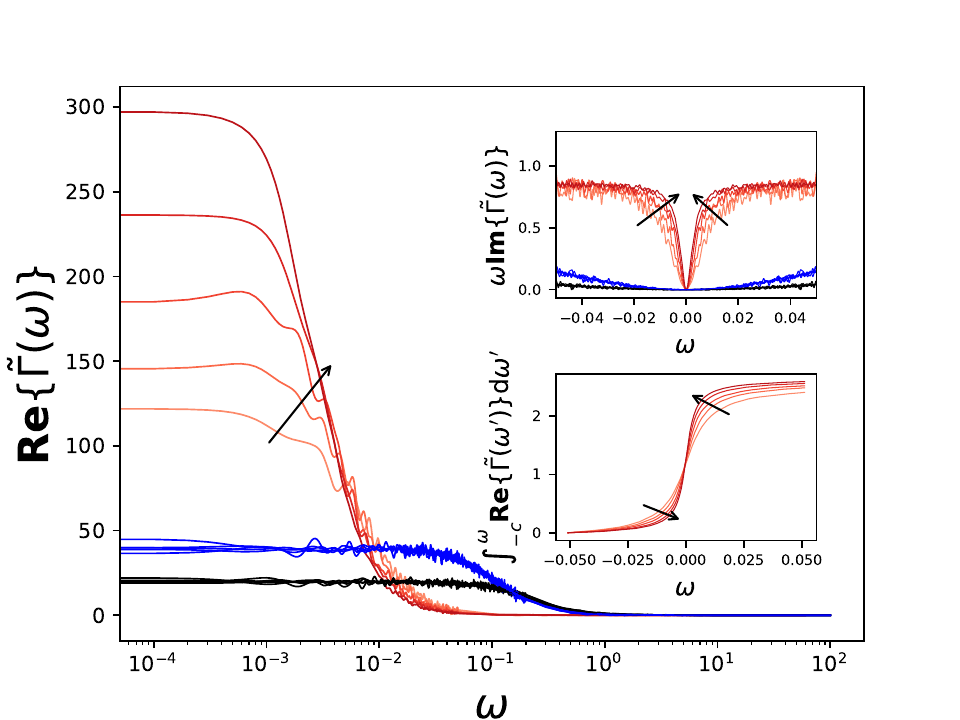}
    \label{fig: subfig 2a}
    }
    \qquad \qquad%\hfill
    \subfloat[]{%
     \includegraphics[trim= 0.215cm 0cm 1.2cm 1.3cm,clip=true, scale=0.5]{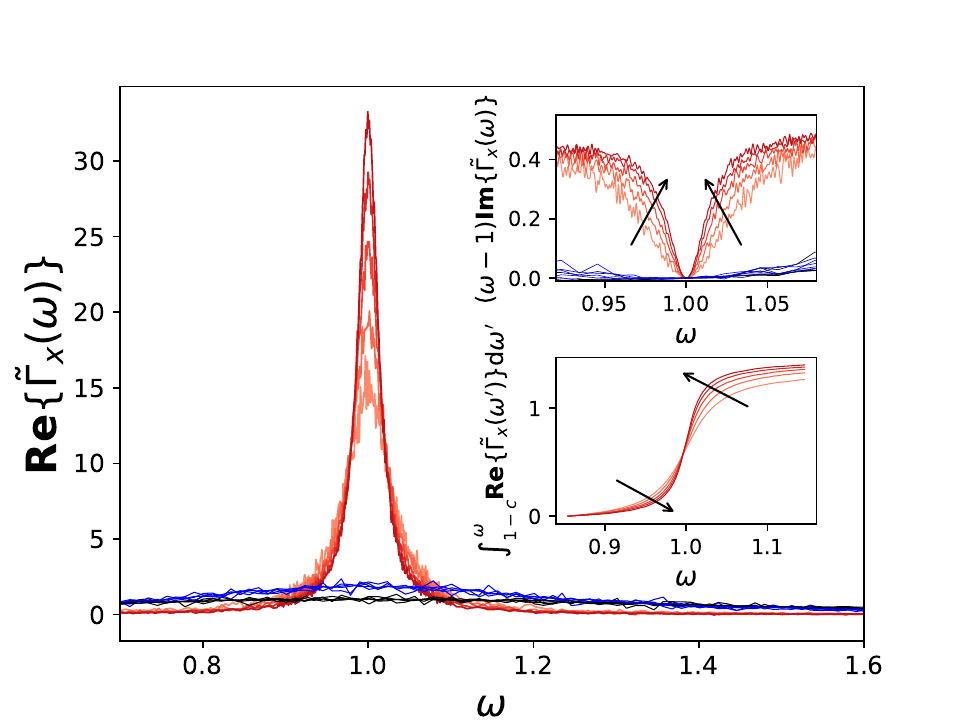} %trim=0.1cm 0.2cm 1cm 1.2cm, clip=true, scale=0.5
      \label{fig: subfig 2b}
    }
    \caption{Macroscopic susceptibilities as a function of the frequency $\omega$.  Panel \protect\subref{fig: subfig 2a}: susceptibility $\tilde{\Gamma}(\omega)$ for the one dimensional order parameter of the DZ model. Panel \protect\subref{fig: subfig 2b}: susceptibility $\tilde{\Gamma}_x(\omega)$ for the first component of the two dimensional order parameter for the BCM model. The parameter in the lower extreme of the integral is for both cases $c=0.05$. Blue and black lines in panel \ref{fig: subfig 2b} have been multiplied by a factor $5$ for visualisation purposes. Colour code and plotting conventions as in Figure \ref{fig: Response in time}.}
\label{fig: Susceptibilities} 
\end{figure*}

As discussed in \cite{FirstPaper}, the critical behaviour of this class of multiagent systems, signified by the singular behaviour of the susceptibility, originates from two distinct physical phenomena that are associated to either the poles of $\Gamma_i(\omega)$ or $P_{ij}^{-1}(\omega)$. The case where $\Gamma_i(\omega)$ diverges pertains to the occurrence of critical transitions.
\\
It is of interest here the case where poles appear in the real $\omega$ axis for $\tilde{\Gamma}_i(\omega)$ because the matrix $P^{-1}_{ij}(\omega)$ becomes singular. 
% that are not poles of $\Gamma_i(\omega)$. 
This corresponds to phase transitions  originated by the coupling and do not show a divergence of correlation properties, because $\Gamma_i(\omega)$ is, instead, analytic in the upper complex $\omega$ plane. Equivalently, the spectral gap of $\mathcal{L}^+_{\langle x \rangle_0}$ remains finite at a phase transition.
However, at the transition point, the usual dispersion relations need to be modified~\cite{FirstPaper,Lucarini2005}.  The conditions underpinning the breakdown of linear response theory do not depend on the perturbation field $\mathbf{X}(\mathbf{x})$ nor on the choice of the observable ($\mathbf{x}$, in our case) and can be related to the spectral properties of a modified transfer operator~\cite{FirstPaper}.

\section{Numerical results\label{sec: numerics}}
Below, we present results for the Desai-Zwanzig (DZ) ~\cite{DesaiZwanzig} and the Bonilla-Casado-Morillo (BCM) ~\cite{Bonilla1987} models. These are composed by $N$ interacting agents evolving according to Eq. \ref{eq1}. Each individual agent of the DZ (BCM) model evolves in $\mathbb{R}$ ($\mathbb{R}^2$). A detailed description of the two models is presented in the Appendix. %We approach the phase transition by keeping fixed the value of the internal parameter $\alpha$ and of the coupling intensity $\Theta$ and by varying, instead, the noise intensity $\sigma$. 
We repeat our experiments for various choices of $N$, in order to detect the emergence of singularities for the combination of the parameters corresponding to phase transitions. Here, we keep fixed the values of the internal parameter $\alpha$. Both models undergo a phase transition at the transition line $\tilde\sigma=\sigma(\theta;\alpha)$ in the  parameter space $(\sigma,\theta)$, see Appendix for the analytical evaluation of the transition line. Since one of the two parameters is redundant, we fix the coupling intensity $\theta$ and we vary, instead, the noise strength $\sigma$. %and of the  The phase transition is obtained for a specific value  that can be calculated analytically for both models, 

Following \cite{Marconi2008}, we perform $n$ simulations where the initial conditions are chosen according to the unperturbed invariant measure $\rho_0(\mathbf{x})$ and where at time $\tau=0$ we apply a perturbation proportional to a Dirac $\delta$ function. The average of the response for the observable $\mathbf{x}$ over the $n$ simulations gives an estimate $\tilde{G}_i(\tau;N)$ of the renormalised response function . Details on the numerical simulations are also reported in the Appendix. %   numerical integration of equations \ref{eq1} for a finite number of particles $N$ and applying a broad frequency perturbation  by delta perturbing the system in time after it has reached its steady state $\rho_0(\mathbf{x})$.  
Figure \ref{fig: Response in time} shows  the response functions $\tilde{G}_i(\tau;N)$ for an additive perturbation $\mathbf{X}(\mathbf{x})=1$ for the DZ model (left panel) and $\mathbf{X}(\mathbf{x}) = (1,0)$ for the BCM model (right panel). The two response functions are qualitatively different because, by and large, the one for the DZ model describes a monotonic decay, whereby the system relaxes towards the unperturbed state, while the one for the BCM combines the decay with an oscillatory behaviour taking place at the natural frequency $\tilde\omega=1$. 

In the DZ model, the response functions initially undergo a fast and substantial decay, both far from and at the phase transition, associated with a time scale of order $1$. % that depends on the profile  $\mathbf{X}(\mathbf{x})$ of the perturbation. 
However, at the phase transition, a new, much longer, timescale appears. This timescale increases monotonically with  $N$. The same is observed in the case of the BCM model if one considers the envelope of the response function rather than the response function itself: at the transition the decay of the oscillations becomes slower and slower as $N$ increases.
\\
The origin of the new timescales resides in the appearance of simple pole at $\omega=\omega_0$ in the susceptibility  $\tilde{\Gamma}_i(\omega;N)$, the Fourier transform of the response function. The pole is located at $\omega_0=0$ for the DZ model and at $\omega_0=\tilde\omega=1$ for the BCM model. When considering finite values of $N$, %we expect that 
the susceptibilities  describing the response of (virtually) any observable to (virtually) any external perturbation have a contribution of the form  %appearance of a simple pole for some $\omega=\omega_0$ for the  leads to the development of the susceptibilities' singular behaviour
$\frac{\kappa}{\omega - \omega_0 + i \gamma(N)}$, where $\gamma(N) \rightarrow 0^+$ as $N \rightarrow +\infty$ and $\kappa$ represents the residue of the pole, because $\lim_{N\rightarrow\infty}\frac{\kappa}{\omega - \omega_0 + i \gamma(N)}=- i\pi\kappa\delta(\omega-\omega_0)+\kappa\mathcal{P}(1/(\omega-\omega_0))$. The quantity $\kappa\in \mathbb{C}$ depends on the choice of observable and of the perturbation. We remark that the asymptotic property does not depend on how fast the function $\gamma(N)$ vanishes for increasing values of $N$. Following \cite{Delgadino2021}, one might conjecture that for the Desai-Zwanzig model and related models the function $\gamma$ would scale as $1/N$. Instead, we have observed here that the behaviour of $\gamma$ is different. This is an issue of fundamental importance that we will explore in future work, also in the case of  nonequilibrium systems. Note also that, in the case of equibrium systems, the mean field limit $N \rightarrow \infty$ and the limit $T \rightarrow T_c$, where $T_c$ is the critical temperature, do not commute \cite{Chavanis2014}.% where we will investigate whether similar conclusions on the $N$-scaling of the spectral gap apply for nonequilibrium systems.}

We next investigate the phase transitions %of the mean field Fokker-Planck equation 
by looking at the properties of the susceptibilities, see Fig. \ref{fig: Susceptibilities}. When $\sigma\neq\tilde\sigma$, the susceptibilities do not show any singularity nor any remarkable dependence on $N$, thus indicating that the thermodynamic limit has been reached to a good approximation. 
As $N$  increases, for both the DZ model (left panel)  and the BCM model (right panel)  the resonance at $\omega = \omega_0$ of the real part of the susceptibility approaches the limiting Dirac function $\pi k\delta(\omega - \omega_0)$ with coefficient $k>0$. This singular behaviour is clear from the plot of the primitive function of the real part of the susceptibility (bottom inset) that tends to step function. % constant times the Heaviside distribution $\Theta(\omega - \omega_0)$. 
For both models $\kappa=\mathrm{i}k$ is an imaginary number. Indeed, the imaginary part of the susceptibility behaves exactly as the Cauchy principal value distribution and can be used to get easily a quantitative estimate of $k$. The top insets of Figure \ref{fig: Susceptibilities} shows the function $(\omega - \omega_0)\mathbf{Im}\{\tilde{\Gamma}_i(\omega)\}$. As $N \rightarrow \infty$, this function converges to $k$ everywhere except for $\omega=\omega_0$.
\\
 An explicit expression for $k$ is known~\cite{FirstPaper} in the case of the DZ model \footnote{There is a typo in the formula given in~\cite{FirstPaper}. Furthermore, the convention for the Fourier transform we  use here has the opposite sign, ditto the residue.}
$ k = \frac{\langle x^2 \rangle_0}{\theta \int_0^{+\infty}\mathrm{d}t\langle x(t)x(0)\rangle_0} $. Using the statistics of the unperturbed runs we obtain $k\approx0.89$, which agrees within 2\% with the one obtained from the limiting behaviour of the susceptibility, thus validating our results. In the case of the BCM model, our procedure allows one to derive a direct estimate $k\approx0.44$; in this case no expression for the residue is available in the literature and, following \cite{Bonilla1987,FirstPaper}, its evaluation seems cumbersome.
We here observe that, by  evaluating the susceptibility for finite values of $N$, we are able to predict the residue of the pole at $\omega=\omega_0$, which appears, instead, only in the thermodynamic limit. The residue plays the role of a latent heat of phase change in classical thermodynamics. Our results, though, allow one to deal with the case of a dynamical latent heat, that is observed for perturbations occurring a non-vanishing frequency. 
As discussed earlier, the singular behaviour of the susceptibility has some degree of universality. By this we mean that while for a given model the value of the residue is forcing- and observable-dependent, its position is a fundamental property of the model itself; 
see  the Appendix for an additional examples. %the study of a spatially dependent perturbation $\mathbf{X}(\mathbf{x}) = \mathbf{x}^2$ for the DZ model.

\section{Conclusions\label{sec: conclusions}}

The study of how a large network of identical agents respond to exogenous perturbations is of the uttermost importance in different areas of science. One might be interested not only in the smooth response of the system, where its properties change ever so slightly, but also in the critical, nonsmooth, regime, where small perturbations can lead to large and possibly undesired changes. Multiagent systems modeled as weakly interacting It\^{o} diffusions represent a rich class of models exhibiting such critical behaviour, and for which rigorous analysis and careful numerical investigations can be carried out. Usually these  phenomena are accompanied by a large spatial (among sub-systems) and temporal restructuring of the system where correlations get highly magnified. The critical behaviour due to the emergence of a phase transition is a genuine thermodynamic phenomenon arising from the complex interactions among the infinite number of agents. Nevertheless, we have shown in this paper that linear response theory provides a powerful framework for detecting and anticipating phase transitions by investigating the response of a finite particle system to external perturbations.%Using methods borrowed from spectroscopy, 
 We have been able to predict the appearance of poles in the susceptibility, which describes the frequency-dependent response of the system, as well as to obtain a correct estimate of critical thermodynamic properties, such as the residue of the poles, based on the knowledge of the response for the finite particle system in two paradigmatic models describing equilibrium and nonequilibrium phase transitions. This is an encouraging starting point for improving our ability to understand and predict transitions in more complex multiagent systems. 
\section*{Data Availability}
The data that support the findings of this study are openly available in "Spectroscopy of Phase Transitions" at \url{https://figshare.com/projects/Spectroscopy_of_phase_transitions/101846}, reference number \cite{Figshare}.
\begin{acknowledgments}
VL acknowledges the support received by the European Union’s Horizon 2020 program through the project TiPES (Grant Agreement No. 820970) and from the EPSRC project EP/T018178/1. The work of GP was partially funded by the EPSRC, grant number EP/P031587/1, and by J.P. Morgan Chase \& Co. Any views or opinions expressed herein are solely those of the authors listed, and may differ from the views and opinions expressed by J.P. Morgan Chase \& Co. or its affiliates. This material is not a product of the Research Department of J.P. Morgan Securities LLC. This material does not constitute a solicitation or offer in any jurisdiction. NZ has been supported by an EPSRC studentship as part of the Centre for Doctoral Training in Mathematics of Planet Earth (grant number EP/L016613/1).
\end{acknowledgments}

\appendix

\begin{section}{The models}
Weakly interacting diffusions represent a rich class of agent based models, describing a network of interacting subsystems. The local dynamics of each subsystem is determined by a smooth vector field $\mathbf{F}_\alpha(\mathbf{x})$. The local force is in general non conservative, leading to irreversible and dissipative processes that can exhibit complex behaviours, such as deterministic chaos, see Figure \ref{fig: weak}.
\begin{figure}
\includegraphics[trim=0.1cm 4.2cm 1cm 5.4cm, clip=true,scale=0.3]{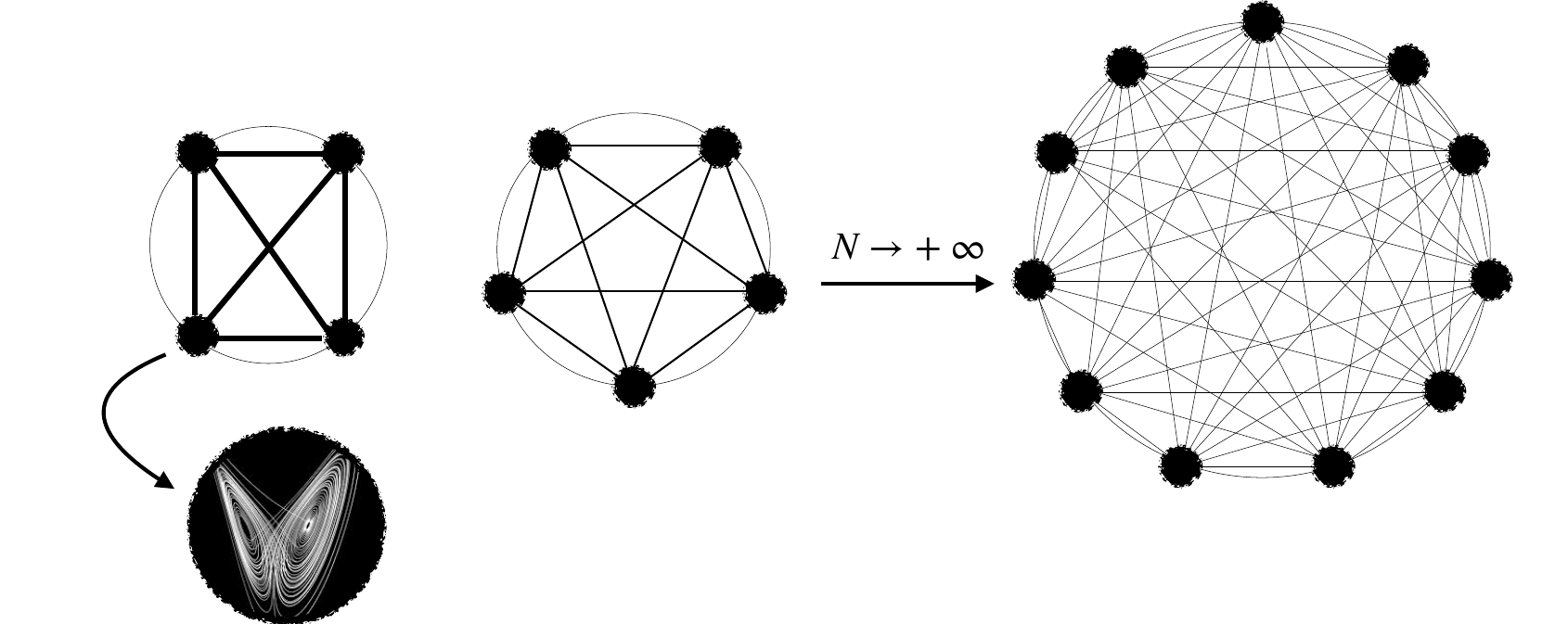}
\caption{Weakly interacting diffusions. The local dynamics of each subsystem, given by $\mathbf{F}_\alpha(\mathbf{x})$, is in general dissipative and can support a wide range of complex behaviours, including deterministic chaos, as depicted in this figure.}
\label{fig: weak}
\end{figure}
\begin{figure*}
    \subfloat[]{%
    \includegraphics[trim= 0.2cm 0cm 1.2cm 1.3cm,clip=true, scale=0.5]{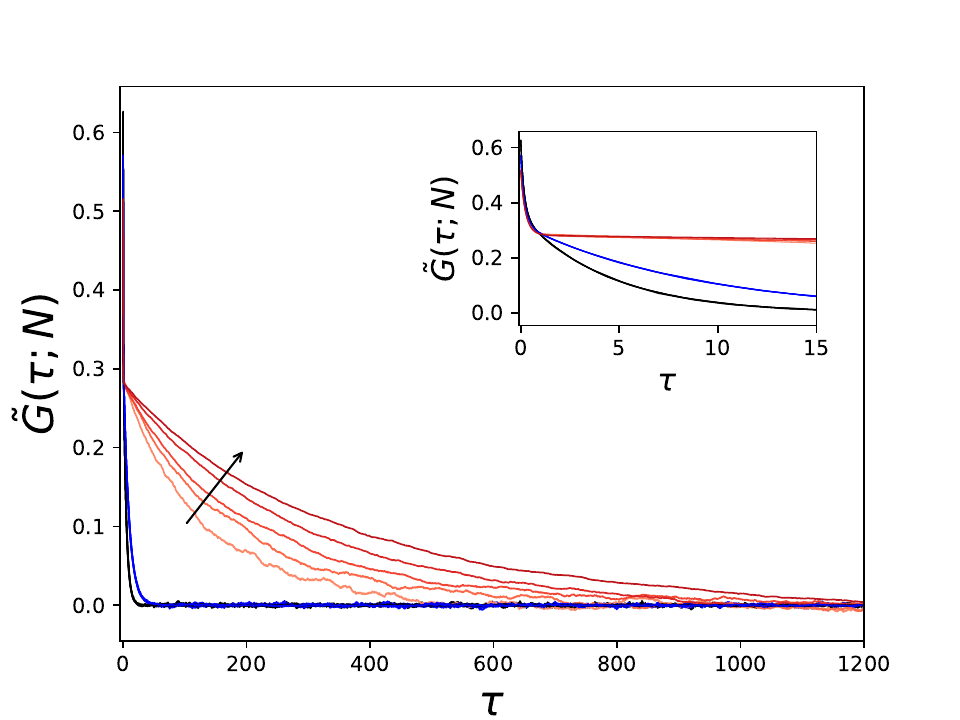}
    \label{fig: subfig 3a}
    }
    \qquad \qquad%\hfill
    \subfloat[]{%
     \includegraphics[trim= 0.2cm 0cm 1.2cm 1.3cm,clip=true, scale=0.5]{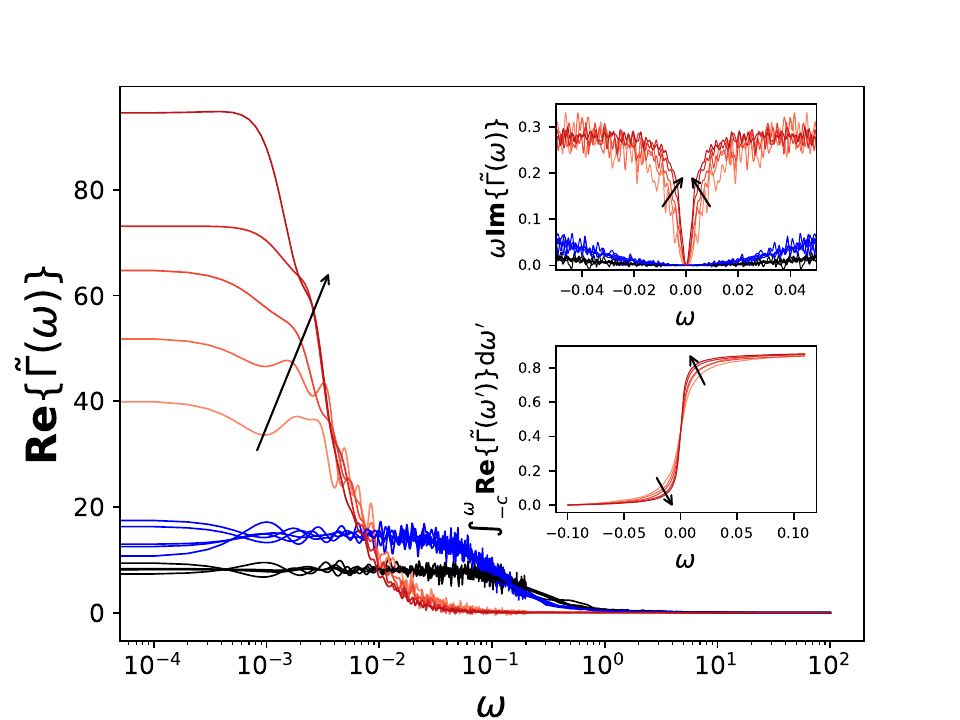} %trim=0.5cm 0.2cm 1cm 1.2cm, clip=true, scale=0.5
     \label{fig: subfig 3b}
     }
     \caption{Panel \protect\subref{fig: subfig 3a}: Response function $\tilde{G}(\tau;N)$ and Panel \protect\subref{fig: subfig 3b}: susceptibility $\tilde{\Gamma}(\omega;N)$  for a spatially dependent perturbation $\mathbf{X}(\mathbf{x})=x^2$ for the one dimensional order parameter for the DZ model. Colour code and plotting conventions as in Figure \ref{fig: Response in time}. The lower extreme of the integral of the bottom panel is $c=0.05$.}
     \label{fig: DZ Mult}
    
\end{figure*}
% \begin{figure*}
%  \includegraphics[scale=0.5]{DesaiRespMult.pdf}
%  \quad 
%  \includegraphics[scale=0.5]{DesaiSuscMult.pdf}
%  \caption{Response functions $\tilde{G}(\tau;N)$ (left) and susceptibility $\tilde{\Gamma}(\omega;N)$ (right) for a spatially dependent perturbation $\mathbf{X}(\mathbf{x})=x^2$ for DZ model. Black and blue lines correspond to non critical values of the strength of the noise $\sigma$. Red lines correspond to response functions at the transition point. For each value of $\sigma$, there are five lines corresponding to different values of $N$, namely $N = 2^k \times 10^3$ with $k = 1, \dots, 5$. The arrows and the colour gradient indicate the direction of increasing $N$. The lower extreme of the integral of the bottom panel is $c=0.05$.}
%  \label{fig: DZ Mult}
% \end{figure*}
An all-to-all coupling between the subsystems is given by a matrix $L_{ij}= \mathbf{\nabla}\mathcal{U}(\mathbf{x}^i - \mathbf{x}^j)$ where $\mathcal{U}(\mathbf{x})= \mathcal{U}(-\mathbf{x})$ represents the interaction potential and $\mathbf{x}^i$ is the state vector of the $i$-th subsystem. Weakly interacting diffusions are characterised by a coupling strength which is inversely proportional to the number of subsystems $N$. As $N$ increases, the interaction structure gets more and more intricate, while the intensity becomes weaker and weaker, see Figure \ref{fig: weak}.
As mentioned in the main text, the DZ model has been introduced, and thereafter commonly used, as a paradigmatic example of an equilibrium continuous phase transition reminiscent of the Ising-like ferromagnetic transitions in spin systems~\cite{Dawson,Shiino1987}. The DZ model describes a network of one-dimensional subsystems $x^k$ whose dynamics is prescribed by the following equations (see main text for notation)
\begin{equation}\label{eq: DZ model}
\mathrm{d}x^k = -V'_\alpha(x^k)\mathrm{d}t -\frac{\theta}{N} \sum_{l=1}^N \left(x^k-x^l\right)\mathrm{d}t+ \sigma\mathrm{d}W^k
\end{equation}
where $k=1,\dots,N$ and the confining potential $V_\alpha(x) = -\frac{\alpha}{2}x^2 + \frac{x^4}{4}$ has a double well shape for $\alpha > 0$. Without loss of generality, we here consider $\alpha =1$. In the absence of coupling, $\theta=0$, the above equations describe the simple motion of a particle in a double well potential, subject to additive noise. The presence of the coupling allows for a long range coordination of the system that in the thermodynamic limit $N \rightarrow +\infty$ results in a proper phase transition. In this regime, by varying the parameters $(\alpha,\theta)$, the order parameter $\langle x \rangle$ undergoes a continuous order-disorder transition, similar to the pitchfork bifurcation diagram for the Ising model. It is possible to show~\cite{Dawson} that the critical line is given by $\frac{D_{-3/2}\left( \frac{\theta - 1}{\sigma}\right)} { D_{-1/2}\left( \frac{\theta - 1}{\sigma}\right)} = \frac{\sigma}{\theta}$
where $D_\nu(z)$ is a parabolic cylinder function. Here the coupling $\theta$ is kept fixed ($\theta = 0.55$) and we vary $\sigma$ to approach the transition point.
 
The BCM model describes an ensemble of bi-dimensional nonlinear oscillators undergoing an out of equilibrium self-synchronisation transition. The time evolution of the network of oscillators is given by the following equations
\begin{equation}\label{eq : Bonilla eqs}
\mathrm{d}\mathbf{x}^{k}=\mathbf{F}_{\alpha}(\mathbf{x}^k) \mathrm{d}t-\frac{\theta}{N}\sum_{l=1}^N \left(\mathbf{x}^k-\mathbf{x}^l\right)\mathrm{d}t+\sigma \mathrm{d}\mathbf{W}^k, 
\end{equation}where the local force is not conservative, giving rise to the non equilibrium features of the system,  and reads $\mathbf{F}_\alpha(\mathbf{x})= \left( \alpha - |\mathbf{x}|^2\right) \mathbf{x}+\mathbf{x}^+ $where $\mathbf{x}^+ = (-x_2,x_1)$. The latter term corresponds to a rotation and makes the stationary state a non equilibrium one. The parameter $\alpha >0$ controls the amplitude of the oscillations of the individual non linear oscillators. In fact, when $\theta=\sigma=0$, each subsystem oscillates as $\mathbf{x}^j(t)=\sqrt{\alpha} \left(\cos(t+\beta_j), \sin(t+\beta_j) \right)$ where $\beta_j = \tan(x^j_2(0)/x^j_1(0))$. The coupling tries to synchronise the subsystems by attracting them towards the center of mass $\frac{1}{N}\sum_{j=1}^N \mathbf{x}^j$. In the thermodynamic limit and for sufficiently low values of the noise, the order parameter $\langle \mathbf{x} \rangle = \langle \mathbf{x} \rangle(t)$ exhibits a periodic time evolution, resulting from the subsystem oscillating in a coherent way. On the other hand, high values of the noise correspond to a non synchronised state where the order parameter vanishes. In particular, the transition happen at the surface of the $(\alpha,\theta,\sigma)$ parametric space defined by~\cite{Bonilla1987}
\begin{equation}\label{eq : transition line Bonilla}
A = \frac{\delta^2}{2}\left[ 1- \frac{1}{\delta}\exp\left(-\frac{A^2}{\delta^2} \right) \left[ \int_{-\frac{A}{\delta}}^\infty e^{-r^2}dr\right]^{-1} \right]  
\end{equation}where $A = \frac{\alpha}{\theta} - 1$ and $\delta = \frac{\sqrt{2\sigma^2}}{\theta}$.
In the following we have $\theta = \alpha = 2$. The colour code for the figures is given by:
\begin{itemize}
\item non critical black : DZ $\sigma \approx 1$ , BCM $\sigma \approx 2$.
\item non critical blue : DZ $\sigma \approx 0.87$ , BCM $\sigma \approx 1.8 $
\item critical red :  DZ $\tilde \sigma \approx 0.75 $ , BCM $ \tilde \sigma \approx 1.59 $.  
\end{itemize}
\subsection{Numerical linear response experiments}
As mentioned in the main text, we perform $n$ simulations of the response given by Eqs. \ref{eq: DZ model} and \ref{eq : Bonilla eqs} where the initial conditions are chosen according to the respective unperturbed invariant measure $\rho_0(\mathbf{x})$ and where at time $\tau=0$ we apply a perturbation proportional to a Dirac's $\delta(\tau = 0)$. The average of the response for the observable $\mathbf{x}$ over the $n$ simulations gives an estimate of $\tilde{G}_i(\tau;N)$. The response functions away from the transitions are estimated on  an ensemble of $n = 10^5$ simulations, while the critical response functions with $n=10^6$ for DZ and $n = 7\times 10^6 $ for BCM. Furthermore we investigate the response up to time $\tau = 5 \times 10^3$. The corresponding susceptibility $\tilde{\Gamma}_i(\omega; N)$ is simply defined as the Fourier Transform of $\tilde{G}_i(\tau;N)$.
In the main text, we show the results for an additive perturbation $\mathbf{X}(\mathbf{x}) \equiv \mathbf{X}$. However, the critical behaviour of the response does not depend on the type of perturbation, modulo a potential degenerate class of perturbations that have zero projection on the invariant measure $\rho_0(\mathbf{x})$. We have thus decided to investigate the response of the DZ model for a spatially dependent perturbation $X(x) = x^2$, see Figure \ref{fig: DZ Mult}. The response function $\tilde{G}_i(\tau;N)$, both away and at the phase transition, has a rapid initial decay with a timescale that is different from the response function shown in the main text. As a matter of fact, the timescale associated to the dominant mode of the response function for $\tau \rightarrow 0$ does in general depend on the applied perturbation~\cite{FirstPaper}. As expected, the response function at the phase transition develops a much longer timescale that increases as the number of particle increases. A more accurate comparison with the result shown in the main text can only be performed in the frequency domain. Figure \ref{fig: DZ Mult} (right panel) shows that, away from the transition, the susceptibilities have a smooth behaviour and no evident dependence on $N$. At the phase transition, the susceptibility develops the expected singular behaviour  $\frac{\kappa}{\omega - \omega_0 + i \gamma(N)}$, where $\gamma(N) \rightarrow 0^+$ as $N \rightarrow +\infty$ due to the appearance of a simple pole $\omega_0=0$. The residue $\kappa$ is purely imaginary and its magnitude $k$ can be inferred by visual inspection of the top inset representing the function$(\omega - \omega_0)\mathbf{Im}\{\tilde \Gamma(\omega)\}$ to be just less than 0.29. As mentioned in the main text, the residue depends both on the observable and on the perturbation $\mathbf{X}(\mathbf{x})$.
\end{section}

\end{document}